\documentclass[prx,superscriptaddress,twocolumn,longbibliography]{revtex4-1}

\usepackage{amsmath}
\usepackage{amssymb}
\usepackage{amsxtra,color}
\usepackage{latexsym}
\usepackage{graphicx}
\usepackage{tikz}
\usepackage{pgfplots}
\usepgfplotslibrary{groupplots}
\pgfplotsset{compat=1.3}
\usepackage{subcaption}
\usepackage{lipsum}
\usepackage{color}

\usepackage{siunitx}

\usepackage[utf8]{inputenc}
\usepackage{MnSymbol}	% use for sumint

\usepackage{hyperref}

\definecolor{MyDarkGreen}{rgb}{0,0.6,0}
\definecolor{MyDarkBlue}{rgb}{0,0,0.8}
\definecolor{MyDarkRed}{rgb}{0.6,0,0.3}

\hypersetup{breaklinks=true, colorlinks=true,plainpages=true, linktocpage=true, linkcolor=MyDarkBlue, citecolor=MyDarkGreen, urlcolor=MyDarkRed, pdfborder={0 0 0},%
pdfauthor={},%
pdfsubject={Research article},
pdftitle={}%
}

\begin{document}
 
\title{A high-repetition rate attosecond light source for time-resolved coincidence spectroscopy}

\author{Sara \surname{Mikaelsson}}
%\email{sara.mikaelsson@fysik.lth.se}
\affiliation{Department of Physics, Lund University, Box 118, SE-221 00 Lund, Sweden}

\author{Jan \surname{Vogelsang}}
\affiliation{Department of Physics, Lund University, Box 118, SE-221 00 Lund, Sweden}

\author{Chen \surname{Guo}}
\affiliation{Department of Physics, Lund University, Box 118, SE-221 00 Lund, Sweden}

\author{Ivan \surname{Sytcevich}}
\affiliation{Department of Physics, Lund University, Box 118, SE-221 00 Lund, Sweden}

\author{Anne-Lise \surname{Viotti}}
\affiliation{Department of Physics, Lund University, Box 118, SE-221 00 Lund, Sweden}

\author{Fabian \surname{Langer}}
\affiliation{Department of Physics, Lund University, Box 118, SE-221 00 Lund, Sweden}

\author{Yu-Chen \surname{Cheng}}
\affiliation{Department of Physics, Lund University, Box 118, SE-221 00 Lund, Sweden}

\author{Saikat \surname{Nandi}}
\affiliation{Department of Physics, Lund University, Box 118, SE-221 00 Lund, Sweden}

\author{Wenjie \surname{Jin}}
\affiliation{ASML Veldhoven, De Run 6501, 5504 DR, Veldhoven, The Netherlands}

\author{Anna \surname{Olofsson}}
\affiliation{Department of Physics, Lund University, Box 118, SE-221 00 Lund, Sweden}

\author{Robin \surname{Weissenbilder}}
\affiliation{Department of Physics, Lund University, Box 118, SE-221 00 Lund, Sweden}

\author{Johan \surname{Mauritsson}}
\affiliation{Department of Physics, Lund University, Box 118, SE-221 00 Lund, Sweden}

\author{Anne \surname{L'Huillier}}
\affiliation{Department of Physics, Lund University, Box 118, SE-221 00 Lund, Sweden}

\author{Mathieu \surname{Gisselbrecht}}
\affiliation{Department of Physics, Lund University, Box 118, SE-221 00 Lund, Sweden}

\author{Cord L. \surname{Arnold}}
\affiliation{Department of Physics, Lund University, Box 118, SE-221 00 Lund, Sweden}

%32.80.-t : Photoionization and excitation
%42.65.Re : Ultrafast processes; optical pulse generation and pulse compression 
%31.15.vj : Electron correlation calculations for atoms and ions: excited states
%32.80.Ee : Rydberg states

\begin{abstract}
Attosecond pulses, produced through high-order harmonic generation in gases, have been successfully used for observing ultrafast, sub-femtosecond electron dynamics in atoms, molecules and solid state systems.
Today's typical attosecond sources, however, are often impaired by their low repetition rate and the resulting insufficient statistics, especially when the number of detectable events per shot is limited. This is the case for experiments where several reaction products must be detected in coincidence, and for surface science applications where space-charge effects compromise spectral and spatial resolution.

In this work, we present an attosecond light source operating at 200\,kHz, which opens up the exploration of phenomena previously inaccessible to attosecond interferometric and spectroscopic techniques. Key to our approach is the combination of a high repetition rate, few-cycle laser source, a specially designed gas target for efficient high harmonic generation, a passively and actively stabilized pump-probe interferometer and an advanced 3D photoelectron/ion momentum detector.
While most experiments in the field of attosecond science so far have been performed with either single attosecond pulses or long trains of pulses, we explore the hitherto mostly overlooked intermediate regime with short trains consisting of only a few attosecond pulses.
We also present the first coincidence measurement of single-photon double ionization of helium with full angular resolution, using an attosecond source. This opens up for future studies of the dynamic evolution of strongly correlated electrons.
\end{abstract}

\maketitle 
\section{Introduction}

The advent of attosecond pulses in the beginning of the millennium \cite{PaulScience2001, HentschelN2001} enabled the study of fundamental light-matter interactions with unprecedented time resolution \cite{KrauszRMP2009}, revealing sub-femtosecond electron dynamics in atoms, molecules and solids, such as ionization time delays \cite{CavalieriNature2007,SchultzeScience2010, KlunderPRL2011, IsingerScience2017}, the change of dielectric polarizability \cite{SchultzeNature2012}, and the timescale of electron correlations \cite{ManssonNP2014, OssianderNatPhys2017}.

Attosecond pulses are generated through high-order harmonic generation (HHG), when intense femtosecond pulses are focused into a generation gas \cite{FerrayJPB1988}. Close to the peak of each half-cycle, an electron wave packet is born through tunnel ionization. It is subsequently accelerated by the electric field of the driving laser pulse and finally may return to its parent ion and recombine, upon which its excess energy is emitted as an attosecond pulse in the extreme ultraviolet (XUV) to soft-X-ray spectral range \cite{CorkumPRL1993, KrausePRL1992}.
The process repeats itself for every half-cycle of the driving field, resulting in an attosecond pulse train (APT) in the time domain and a comb of odd-order harmonics in the frequency domain.
If the emission originates from only one half-cycle, a single attosecond pulse (SAP) is emitted with a continuous frequency spectrum.

Two well-established pump-probe techniques, based on cross-correlating the attosecond pulses with a low frequency field (usually a replica of the generating pulse) while the photoelectron spectrum originating from a detection gas is recorded, give access to dynamics on the attosecond time scale.  
The RABBIT (Reconstruction of Attosecond harmonic Beating By Interference of Two-photon transitions) is well suited for the characterization and use of APTs, while the streaking technique is mostly applied to SAPs \cite{PaulScience2001, HentschelN2001}.
The requirements to perform such experiments are challenging in terms of laser sources, HHG and pump-probe interferometric optical setups, as well as photoelectron detectors.
Traditionally, mostly chirped pulse amplification, Titanium:Sapphire-based lasers with repetition rates in the low kHz range have been used, rendering experiments that have high demands on statistics or signal-to-noise ratio (SNR) time consuming \cite{ManssonNP2014}.

Here, we present a high-repetition rate, flexible attosecond light source, particularly designed for the study of gas phase correlated electron dynamics as well as time-resolved nano-scale imaging. This article both summarizes and extends previous work \cite{HarthJO2017,ChengPNAS2020}. 
The laser system, located at the Lund High-Power Facility of the Lund Laser Centre, is based on optical parametric chirped pulse amplification (OPCPA), providing sub-\SI{6}{fs} long pulses with stabilized carrier-to-envelope phase (CEP) in the near-infrared with up to \SI{15}{\micro\joule} pulse energy at a repetition rate of \SI{200}{\kilo\hertz} \cite{HarthJO2017}.
The 200-fold increase in repetition rate, compared to standard \SI{1}{\kilo\hertz} systems, promotes experiments with high demands on statistics. A 3D momentum spectrometer \cite{GisselbrechtRSI2005}, capable of measuring several correlated photoelectrons/ions in coincidence, fully resolved in momentum and emission direction, has been installed as a permanent experimental end station.
The combination of CEP control, where the phase of the electric field is locked to the pulse envelope, and the short pulses, comprising only two cycles of the carrier wavelength, provides control of the characteristics of the generated APTs. For example, we can choose the number of pulses in the train to be equal to two or three pulses, which allows exploring the transition between the traditional streaking and RABBIT regimes. As a proof of concept for a statistically very demanding experiment, we present measurements on single-photon double ionization of helium, which is an archetype system of strongly correlated pairs of photoelectrons \cite{ChandraJPB2002,ChandraPRA2004,AkouryScience2007}.

\begin{figure}[ht!]
\includegraphics[width=1\linewidth]{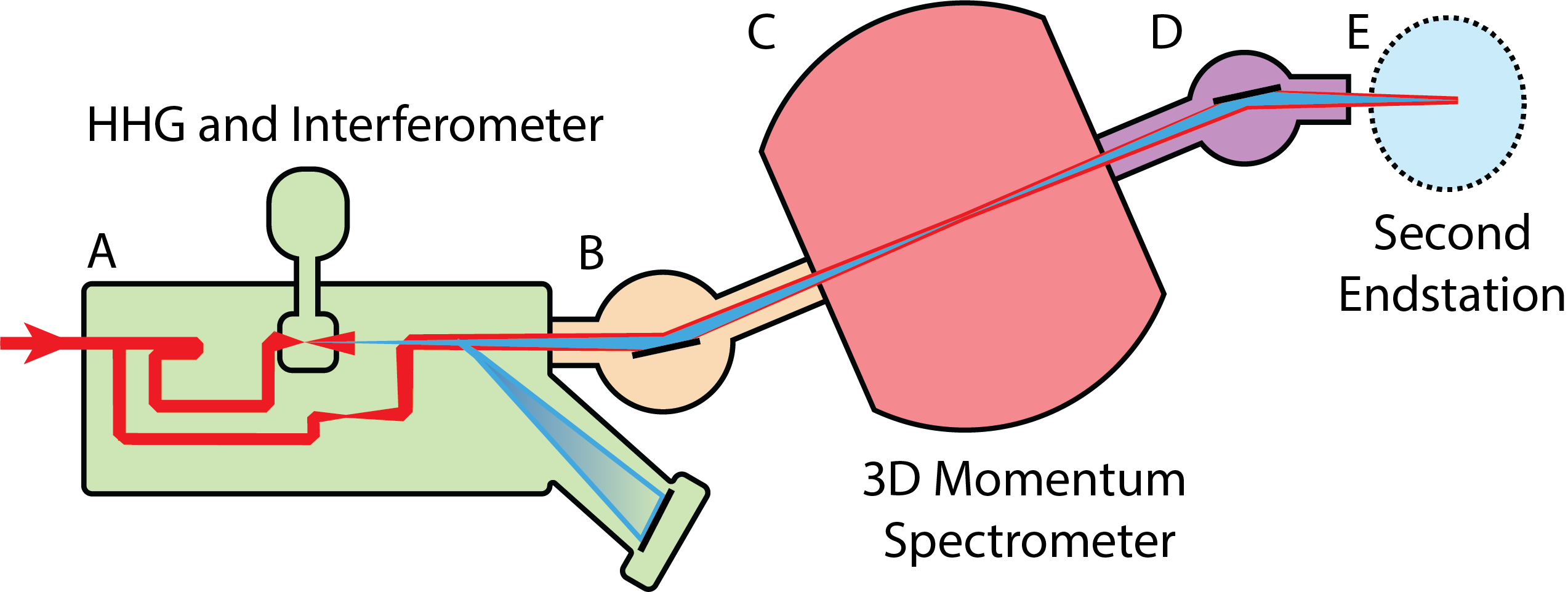}%
\caption{\textbf{Beamline footprint}. (A) HHG generation and characterization and XUV-IR interferometer. (B) Focusing chamber. (C) First interaction region, 3D momentum spectrometer. (D) Refocusing chamber. (E) Second interaction region, flexible endstation. \label{Fig:beamline}}%
\end{figure}

Figure \ref{Fig:beamline} shows a schematic of the experimental setup. The laser pulses enter the first vacuum chamber (A), green, which contains a pump-probe interferometer, a small gas jet for HHG and an XUV spectrometer for diagnostics. The XUV pump and infrared (IR) probe are then focused by a toroidal mirror (B), yellow, into the sensitive region of the 3D momentum spectrometer (C), red. A refocusing chamber (D), purple, contains a second toroidal mirror for re-imaging to the second interaction region, (E), where different end stations for surface science, usually a photoemission electron microscope (PEEM), can be installed \cite{StockmanNP2007,Chew2015,ZhongNC2020}. The beamline is designed for simultaneous operation of both end stations.

The paper is structured as follows: The first section introduces the optical setup by briefly discussing the laser source and the XUV-IR pump-probe interferometer, before examining the gas target design for HHG and the control of the emitted attosecond pulse trains and finally, introducing the 3D photoelectron/ion spectrometer. The following section discuss the utilization of our light source for attosecond time-resolved spectroscopy. We close by presenting measurements of the fully differential cross-section for double ionization of helium, an experiment that to our best knowledge has not been performed with attosecond pulses before.

% -----------------------------------------------------------------------
\section{XUV Light source and pump-probe setup}
\subsection{Laser Source Characterization}
\begin{figure}[ht!]
\includegraphics[width=1\linewidth]{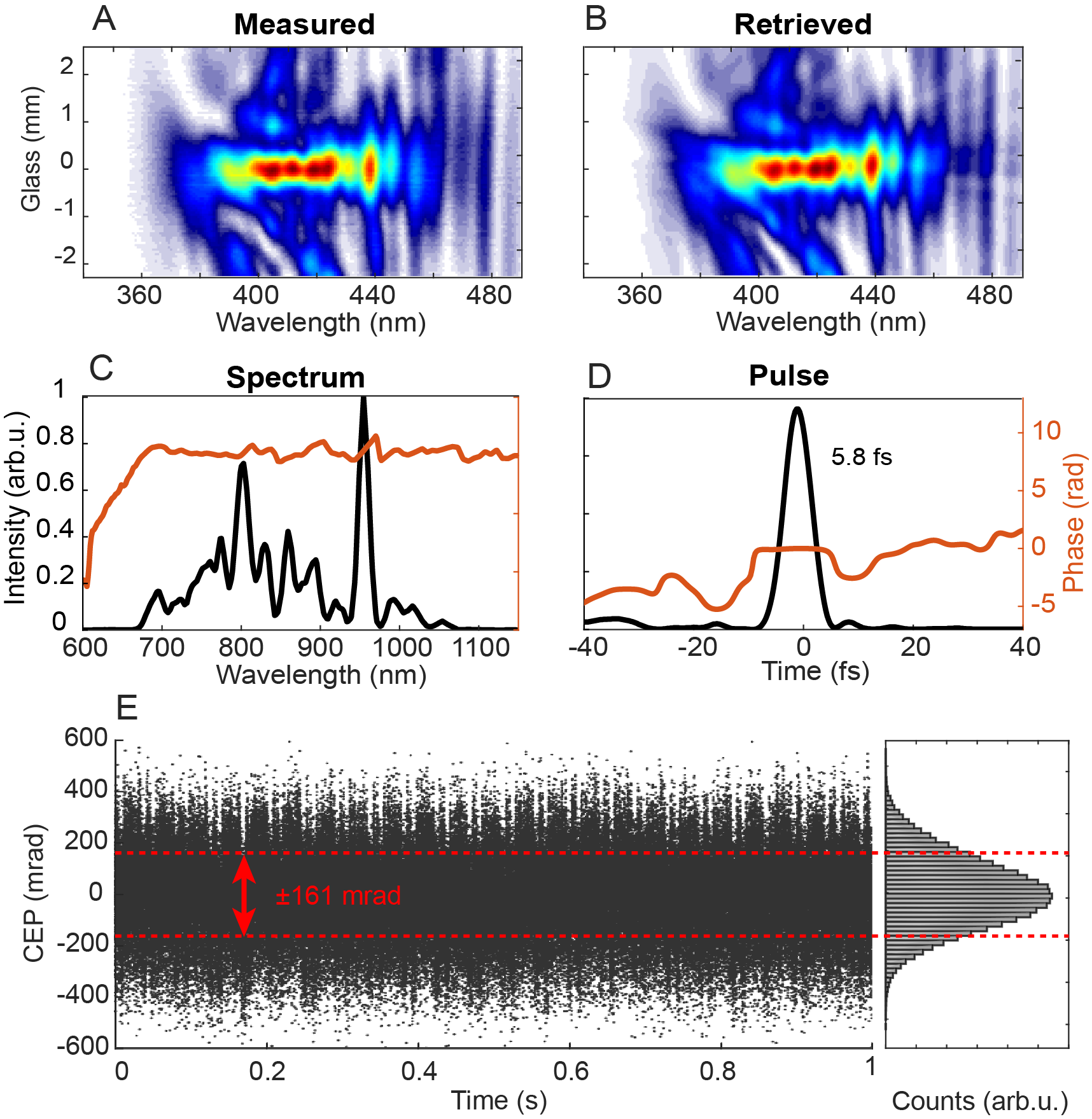}%
\caption{\textbf{Laser output characterization.} (A) Measured and (B) retrieved d-scan traces; a d-scan trace is a two-dimensional representation of the second harmonic signal (intensity in colors) as a function of wavelength and glass insertion; (C) Retrieved spectrum (black) and spectral phase (red), (D) Retrieved temporal pulse intensity profile (black) and phase (red), (E) Single-shot CEP measurement, showing the measured CEP as a function of time; The root mean square is indicated in red (see statistics on the right). \label{Fig:dscan}}%
\end{figure}

The OPCPA laser that the beamline is operated with is seeded by a Titanium:Sapphire ultrafast oscillator. The seed pulses are amplified in two non-collinear optical parametric amplification stages, pumped by a frequency-doubled, optically synchronized Ytterbium-fiber chirped pulse amplifier (CPA). The details of the laser are described elsewhere \cite{HarthJO2017}. 
Figure \ref{Fig:dscan} shows temporal characterization of the output pulses performed with the dispersion scan (d-scan) technique \cite{MirandaOE2012a}, revealing a pulse duration of \SI{5.8}{fs} full width at half maximum (FWHM) at a carrier wavelength in the near-infrared of approximately 850\,nm.

The CEP stability of the laser was measured out-of-loop with a f:2f interferometer \cite{TelleAPB1999}, capable of single-shot acquisition at full repetition rate. This is achieved by recording the spectral interference, encoding the CEP signal, with a high-speed line camera (Basler) with an acquisition rate of $>$\SI{200}{\kilo\hertz}.
Figure \ref{Fig:dscan}E shows measurement results demonstrating a short term CEP stability with a Root Mean Square (RMS) of approximately \SI{160}{\milli\radian}. 
Single-shot CEP detection is an essential tool to ensure that our light source is stable. Furthermore, the CEP data can be tagged to additionally reduce the noise in photoelectron/ion data.
An additional f:2f interferometer is used to actively compensate for long term drift.

% -----------------------------------------------------------------------
\subsection{XUV-IR Interferometer}

\begin{figure*}[ht!]
\includegraphics[width=1\linewidth]{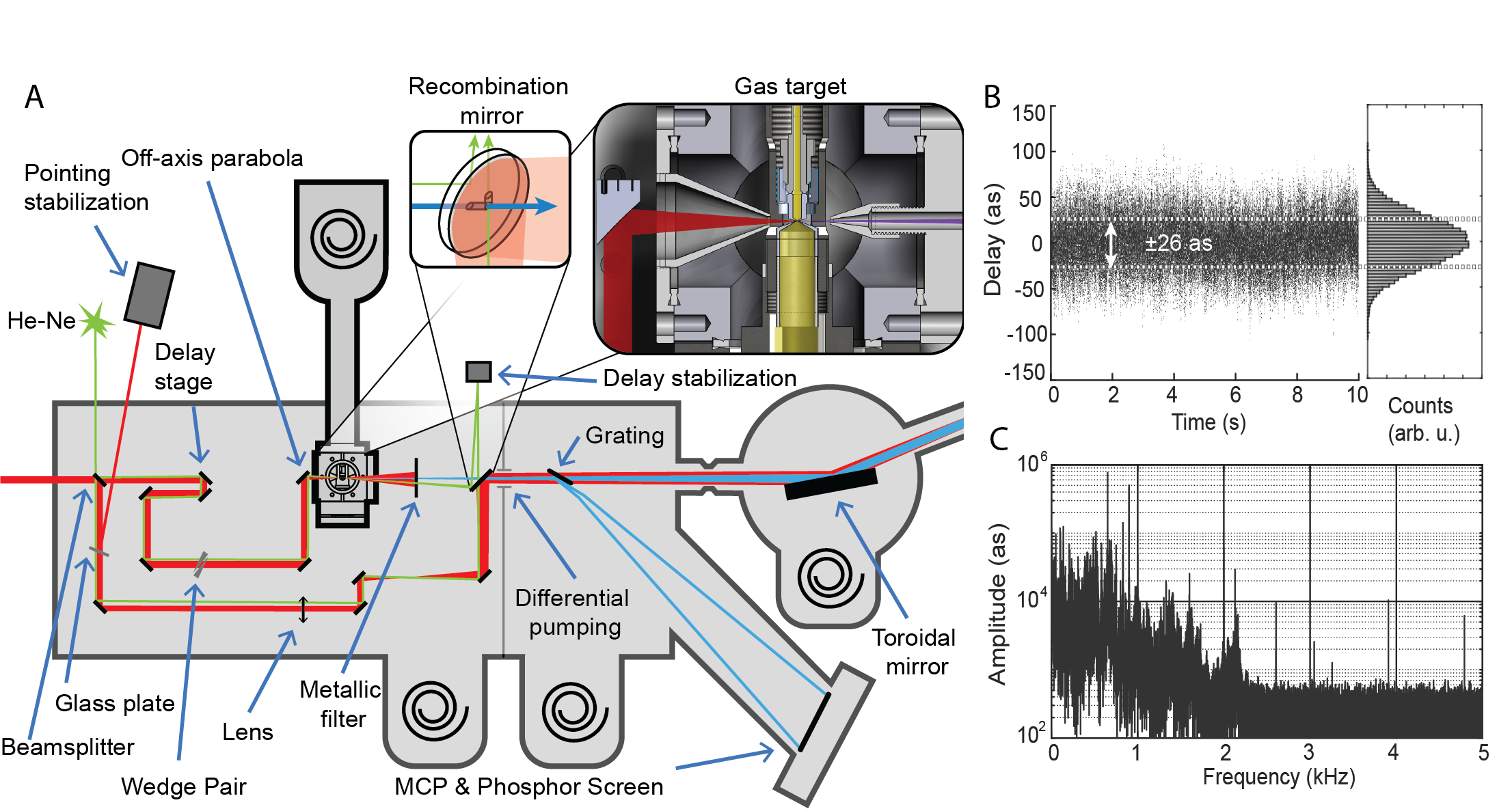}%
\caption{\textbf{XUV-IR interferometer.} (A) Interferometer beam-path and components. The near-IR beam path is shown in red, the XUV in blue, and a HeNe-laser which can be used for active delay stabilization is shown in green. (B) Stability measurement, showing the measured delay as a function of time; The root mean square is indicated in white (see statistics on the right). (C) Single sided amplitude spectrum of interferometer stability measurement. \label{Fig:Infr}}%
\end{figure*}

The XUV-IR pump-probe interferometer is designed for high temporal stability. Suppressing instabilities from vibrations and drifts as much as possible is essential for recording data with high SNR \cite{IsingerPTRSA2019} over several hours or even days. To promote passive mechanical stability, all optical components of the interferometer are mounted on a single mechanically stable breadboard inside the main vacuum chamber. The breadboard is stiffly mounted to an optical table, but vibrationally decoupled from the vacuum chamber in order to isolate from vibrations originating from turbo and roughing pumps. Figure \ref{Fig:Infr}A shows the interferometer layout and beam path. 

After entering the vacuum chamber, the laser pulses are split into pump- and probe arms of the interferometer by a thin beamsplitter optimized for ultrashort laser pulses (Thorlabs UFBS2080), reserving 80\% of the power for HHG in the so-called pump arm and 20\% for the probe arm. In view of the short pulse duration and broad spectrum of the few-cycle pulses, dispersion management between the two interferometer arms is imperative. It is achieved by a 1\,mm thick AR-coated glass plate in the probe arm (compensating the beam-splitter) and a pair of fused silica wedges at Brewster angle in the pump arm (for other dispersive elements).
A spurious reflex from the glass plate in the probe arm is used to feed a beam pointing stabilization system (TEM-Messtechnik) outside the vacuum chamber.

The delay between the interferometer arms is controlled by a retro reflector in the pump arm, mounted on a linear piezo stage (Piezo Jena Systems) with \SI{80}{\micro\meter} travel.
The pump arm is then focused by a \SI{90}{\degree}, off-axis parabola into the generation gas target for HHG, shown in an insert in Figure \ref{Fig:Infr}A and described in more detail in the next section.
In the probe arm, a lens is used to mimic the focusing in the pump arm, ensuring that after recombination pump- and probe pulses will focus at the same position.
In the pump arm, the generated XUV attosecond pulses are separated from the driving near-infrared pulses by different thin (usually \SI{200}{\nm}) metallic filters, transparent to the XUV but opaque to the driving pulses. In order to reduce the thermal load on the filter, an aperture cutting most of the driving pulses is placed before the filter, utilizing the smaller divergence of the attosecond pulses.
A holey mirror at \SI{45}{\degree} is used for recombining the probe- and pump arms of the interferometer (see recombination mirror inset in Figure \ref{Fig:Infr}A). The XUV attosecond pulses pass through a \SI{1.6}{\milli\meter} wide hole, chosen such that the XUV beam is barely clipped, while the light from the probe arm is reflected as a ring by the holey surface.
Afterwards, a toroidal mirror is used to image the foci of the pump- and probe interferometer arms in 2f:2f configuration into the sensitive region of the 3D photoelectron spectrometer placed downstream.
 The focal length of the lens in the probe arm of the interferometer is chosen to homogeneously illuminate the toroidal mirror, at 78\si{\degree} angle of incidence with a clear aperture of 125$\times$25\,mm and 350\,mm focal length, and thus maximize the pulse energy that is refocused. 

In order to stabilize the delay for slow thermal drift, we couple a Helium-Neon (He-Ne) laser beam into the interferometer, shown in green in Figure \ref{Fig:Infr}A. This is achieved by illuminating the beamsplitter from the so far unused side. The metallic filter after HHG is suspended on a transparent substrate to transmit the part of the HeNe-laser beam with large divergence, which is then reflected by the back-side-polished recombination mirror (see recombination mirror inset of Figure \ref{Fig:Infr}A). Light from the other arm of the interferometer is transmitted through another hole in the recombination mirror. A spatial interference pattern is observed with a camera and used to record the phase drift in order to feed back to the delay stage.
Figure~\ref{Fig:Infr}B shows the short term stability of the interferometer, determined by using a fast CCD detector (10 kHz read-out). This fast detection enables us to get information on high  acoustic frequencies that might contribute to instabilities (Figure~\ref{Fig:Infr}C). 
The sharp peaks in Figure~\ref{Fig:Infr}C can be attributed to the different turbo pumps.
A short-term stability of \SI{26}{\atto\second} is achieved.

% -----------------------------------------------------------------------
\subsection{Gas Target}

As a consequence of the high repetition rate of the few-cycle laser system, the energy of the individual pulses is rather low, compared to the kHz repetition rate Titanium:Sapphire lasers often used for generating high-order harmonics.
This has important implications for HHG: First, tight focusing is required to achieve sufficiently high intensity and second,  phase-matching and scaling considerations imply a localized, high density generation gas target \cite{HeylJPBAMOP2017,HeylO2016a}.  
The off-axis parabola used to focus the driving pulses into the gas target has a focal length of $f=\SI{5}{\cm}$, resulting in a focal radius around \SI{5}{\micro\meter} and intensities in excess of several $10^{14}$\,\si{\watt/\cm^2} can easily be achieved. 
Numerical simulations, based on solving the time dependent Schrödinger equation for the nonlinear response and the wave equation for propagation \cite{LhuillierPRA1992,WikmarkPNAS2018}, suggest a gas density corresponding to ~5 bar pressure and a medium length of ~40 µm. 
The gas density should fall off as rapidly as possible outside the interaction region in order to avoid re-absorption of the generated XUV radiation. Such a gas target imposes substantial engineering challenges.

The gas target consists of a nozzle with an exit hole of 42\,µm, operated for the case of argon at 12\,bars of backing pressure. The small nozzle exit diameter ensures longitudinal (i.e. along the laser propagation direction) and transverse confinement of the interaction region.
However, the high mass flow rate, around $4\times10^{-3}$\,g/s, of such a nozzle, if used to inject gas directly into the vacuum, would challenge the capacity of the turbo pumps and contaminate the vacuum inside the chamber, resulting in severe re-absorption. 
Therefore, we use a catcher with a hole of \SI{1}{\mm} diameter mounted less than \SI{200}{\um} from the injection nozzle and connected to a separate roughing pump. The goal is to catch the majority of the injected gas directly before it expands into the vacuum.
The nozzle-catcher configuration is attached to a manual three dimensional translation stage, for adjusting the position of the gas target with respect to the laser focus without changing the positions of nozzle and catcher with respect to each other.
Comparing the pressure inside the vacuum chamber with and without the catcher indicates that the catcher takes away $>$90\% of the injected gas.

In order to further reduce possible contamination of the surrounding vacuum, the gas target is placed in an additional small chamber, pumped by an extra turbo pump (see gas target inset in Figure \ref{Fig:Infr}A). The laser enters and exits the cube-shaped chamber through differential pumping holes, placed as closely to the interaction region as possible.
As a result, the beam only traverses a distance of approximately \SI{3}{\mm} inside this comparably high pressure environment. During regular operation, when generating in argon (\SI{12}{bar} backing pressure), the pressure inside the cube is around \SI{3e-3}{\milli\bar}, while the pressure in the surrounding chamber remains at \SI{2.5e-5}{\milli\bar}. Following the beam path towards the first experimental chamber, the pressure further reduces via differential pumping to \SI{1e-6}{\milli\bar} in the XUV spectrometer compartment, to \SI{2e-8}{\milli\bar} in the chamber housing the toroidal mirror and finally to \SI{2e-9}{\milli\bar} at the 3D photoelectron spectrometer. A pressure below \SI{5e-10}{\milli\bar} in the second end station is easily reached. 

\begin{figure}[ht!]
\includegraphics[width=1\linewidth]{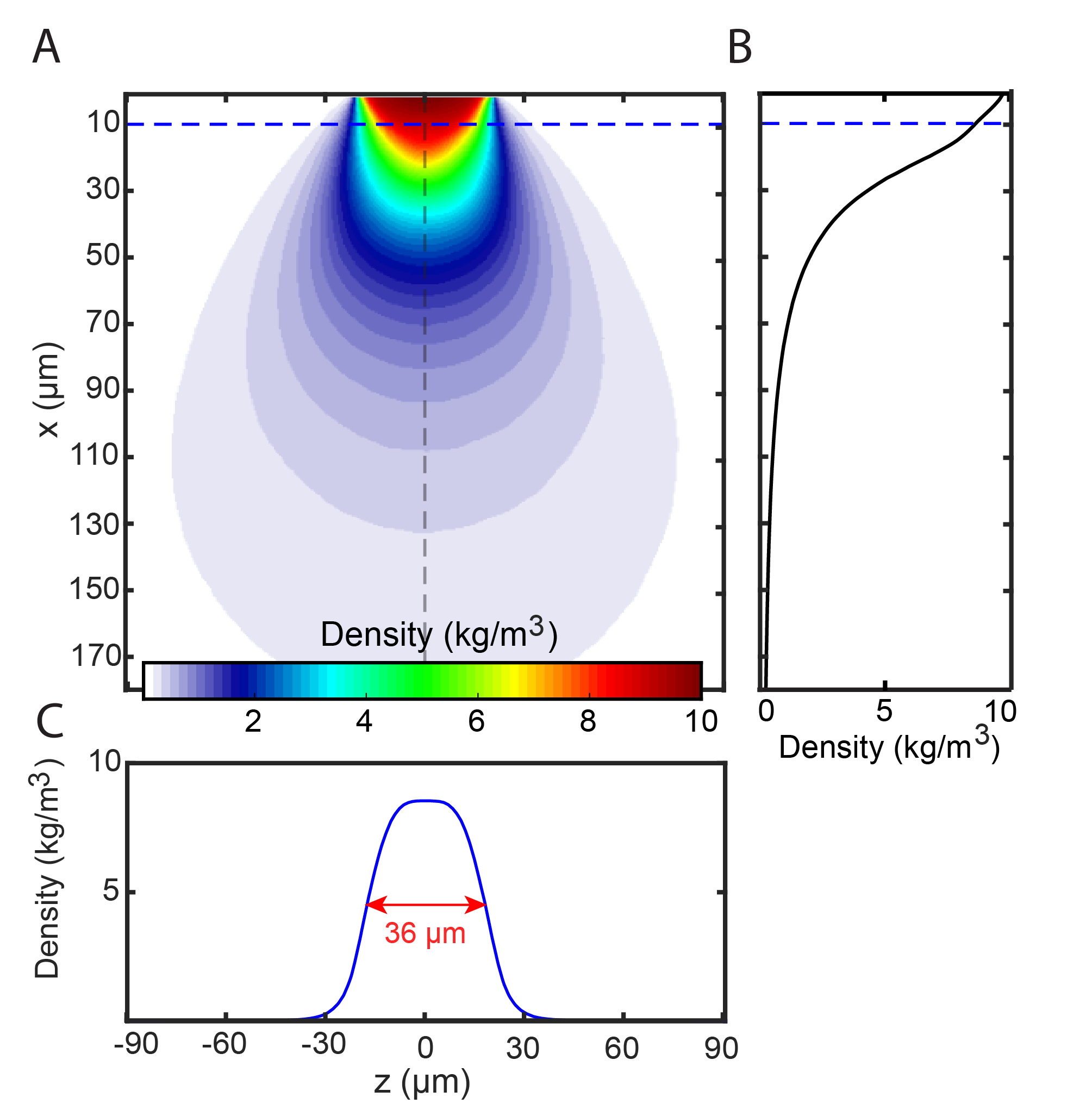}%
\caption{\textbf{HHG gas target simulations.} (A) Simulated gas density in the interaction region with a \SI{42}{\um} nozzle centered around z = 0. (B) Density at z\,=\,0 (black). (C) Line out of the gas density at a distance of 10~\si{\um} from the nozzle. In A and B the dashed blue line indicates the position of the line-out in C. In A the dashed grey line indicates the position of the line-out in B. \label{Fig:Nozzle}}%
\end{figure}

Simulations of the gas density in the interaction region were performed using the STARCCM+ compressible flow solver. The results are summarized in Figure \ref{Fig:Nozzle}. The line out along the direction of the gas flow (Figure \ref{Fig:Nozzle}B) indicates that the density drops rapidly away from the nozzle. 
Figure \ref{Fig:Nozzle}C shows a line out transversely to the gas flow at a distance of 10\,µm from the nozzle exit (marked with a dashed blue line), which corresponds to the approximate distance at which the laser traverses the gas target during operation. According to the simulations, the gas target has an approximately super-Gaussian shape with a width of 36\,µm (FWHM) at this distance to the nozzle with a peak density of \SI{8.5}{\kilogram/\meter^3}, corresponding to 4.7 times the density of argon at standard pressure and temperature condition (1.7\,kg/m$^3$).
The obtained density and gas medium length are in excellent agreement with the calculated best phase matching conditions (5 bar, 40\,µm). Experimentally, we used the fringe pattern resulting from the interference of the stabilization He-Ne laser pump and probe beams (see Fig.~\ref{Fig:Infr}A) in order to get information on the actual gas density.
We measured a phase shift of the fringes, between the cases of active and inactive gas target, equal to \SI{0.65}{\radian}.
Using refractive index data for argon \cite{BorzsonyiAO2008}, a phase shift of \SI{0.625}{\radian} can be calculated from the simulation data in \ref{Fig:Nozzle}C, indicating that the simulation results are in close match to the actual conditions in our gas target.

% ----------------------------------------------------------------------- 

\subsection{High-Order Harmonic Generation}

The process of HHG significantly changes from the case of multi-cycle, long driving pulses, where the attosecond pulses emitted from subsequent half-cycles are nearly identical except for a $\pi$-phase shift between them, to the case of few-cycle driving pulses, where attosecond pulses emitted from consecutive half-cycles can be very different from each other.
In this case, attosecond pulses are only emitted by the most intense half-cycles with an amplitude that is nonlinearly related to the field strength during the half-cycle, the delay between them is not strictly the time between two half-cycles and their phase difference is not exactly $\pi$ \cite{GuoJPB2018}. These properties are also reflected in the spectrum, with spectral peaks which are not necessarily located at the expected positions for odd-order harmonics.

The electric field of the few-cycle pulse, and consequently the characteristics of the HHG pulse train strongly depend on the CEP of the driving pulses, implying that a stable laser CEP is required for generating a reproducible HHG spectrum and attosecond pulse train~\cite{BaltuskaNature2003}. 
For the two-cycle, sub-6\,fs pulses provided by our OPCPA laser, the APTs consist of either two or three strong attosecond pulses, when the CEP of the driving pulse is $\pi/2$ (sine-like pulse) and 0 (cosine-like pulse), respectively \cite{ChengPNAS2020}.
The CEP is controlled by tuning the dispersion of the pulses with a combination of negatively chirped mirrors and motorized BK7 wedges. 

\begin{figure}[ht!]
\includegraphics[width=1\linewidth]{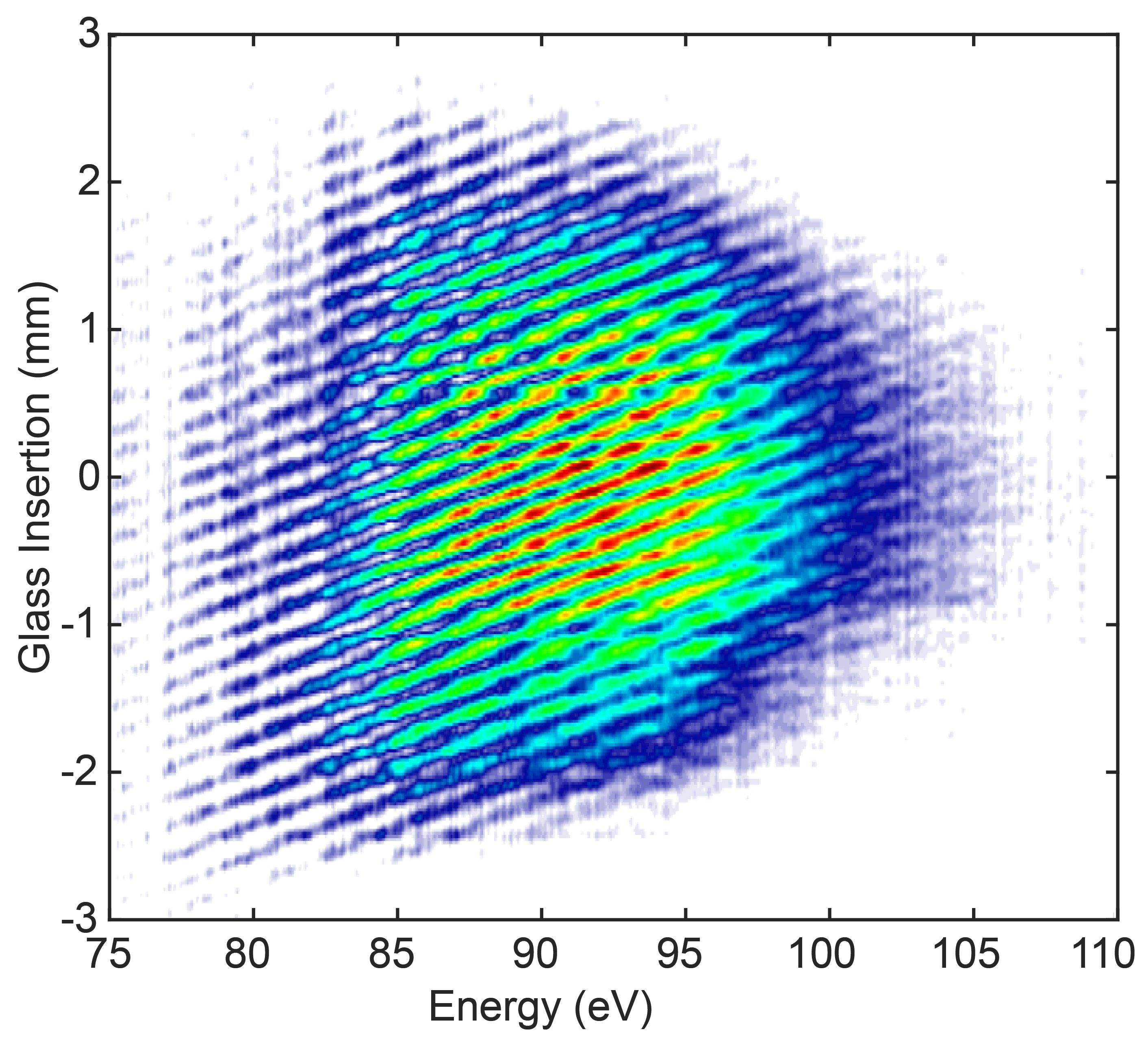}%
\caption{\textbf{HHG Dispersion Scan.} Spectrogram showing the XUV spectrum (horizontal scale) as a function of BK7 glass insertion for HHG in neon. The spectrum is filtered by a 200~\si{\nano\meter} thick zirconium thin film. \label{Fig:DHHG}}%
\end{figure}

To demonstrate the pulse train control achieved in our experiments, we present in Figure~\ref{Fig:DHHG} the XUV spectrum obtained using Ne gas for the generation, as a function of glass insertion. A zirconium filter, cutting all emission below $ 65$\,eV, was used to record the spectrum. For large glass insertion (i.e. $>$~2.5~\si{\mm} and $<$~-2.5~\si{\mm}), the pulse is stretched in time to the extent that the peak intensity becomes too low for HHG, while for small amounts of dispersion, the effect is primarily a change of the CEP. As the CEP varies, the position of the harmonics shift approximately linearly over the whole dispersion and energy range in Figure~\ref{Fig:DHHG}, reflecting a CEP-dependent change of the relative phase between consecutive attosecond pulses. This originates from the CEP-dependent variation of the strength of the field oscillations used for the attosecond pulse generation \cite{GuoJPB2018}.
The highest photon energies with up to \SI{100}{\electronvolt} are obtained at zero glass insertion, i.e. for the shortest pulses and the highest intensity.
The photon flux was characterized in an earlier stage of the light source development (see \cite{HarthJO2017} for details). We estimate the current fluxes to be higher than $15 \times 10^{10}$\,photons/s in argon and $0.7 \times 10^{10}$\,photons/s in neon, respectively.

% -----------------------------------------------------------------------
\subsection{3D photoelectron/ion spectrometer}

\begin{figure}[ht!]
\includegraphics[width=1\linewidth]{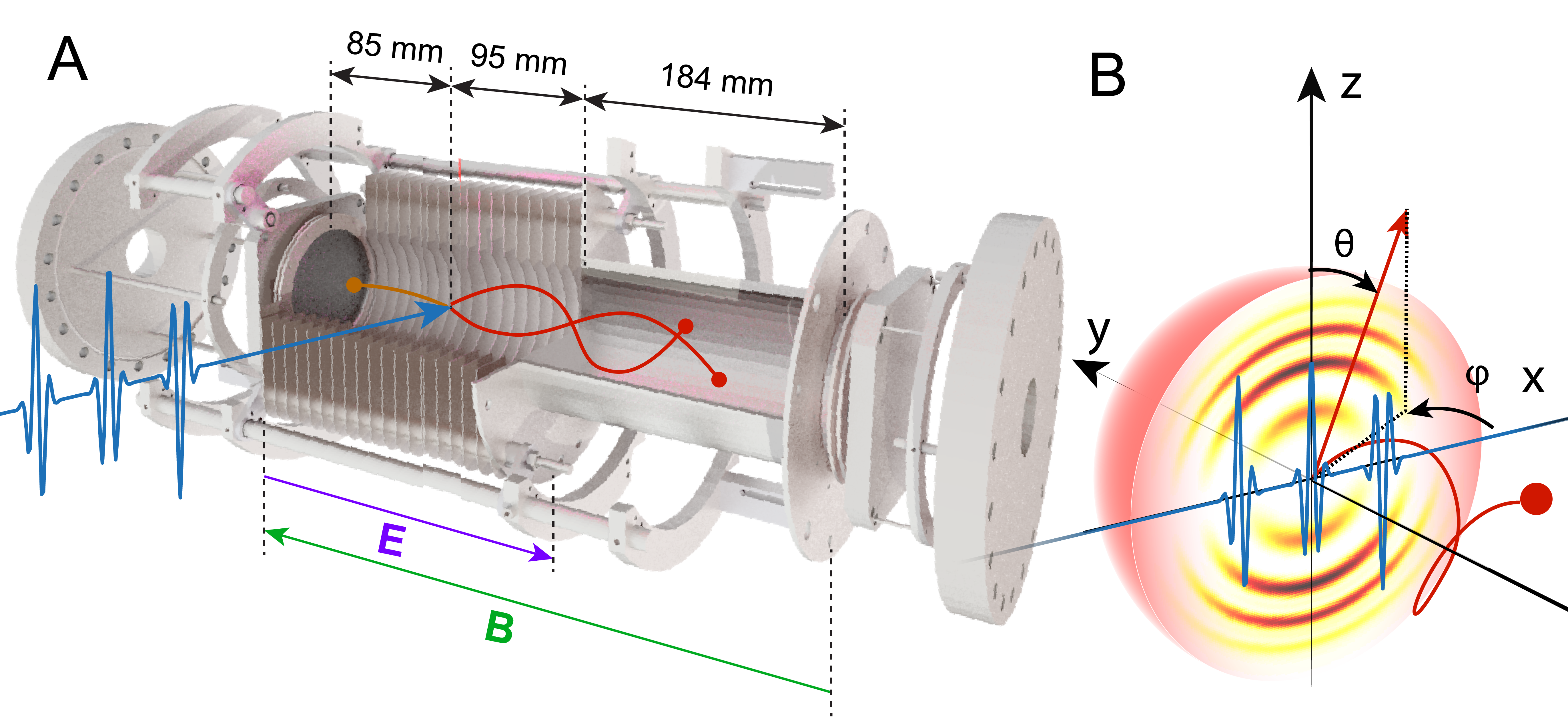}%
\caption{\textbf{Reaction microscope.} (A) Working principle of a reaction microscope. (B) Coordinate system used for measurements, along with a 3D representation of the electron momentum distribution, indicating the azimutal angle ($\varphi$) and elevation angle ($\theta$) with respect to the plane of the laser polarization (along the $z$ axis).\label{Fig:CIEL}}%
\end{figure}

A schematic overview of the 3D photoelectron/ion spectrometer used to study attosecond dynamics in atoms or molecules in gas phase is shown in Figure \ref{Fig:CIEL}. This spectrometer is based on a revised CIEL (Coincidences entre ions et electrons localisés) design \cite{GisselbrechtRSI2005}, which is conceptually similar to REMI (reaction microscope) or COLTRIMS (Cold Target Recoil Ion Momentum Spectroscopy) \cite{DornerPhysRep2000,Ullrich03rpp}.
Momentum imaging instruments of this type have been widely used for the study of photoionization dynamics \cite{HoglePRL2015,HeuserPRA2016} and can with the help of electric and magnetic fields, collect high-energy electrons over the full solid angle (i.e. $4\pi$ collection). The charged particles produced through ionization are accelerated with a weak electric extraction field. In order for the lighter electrons not to escape the spectrometer before they reach the detector, a magnetic field is applied over the whole spectrometer, which confines the electrons to a periodic cyclotron motion with a radius determined by their momentum and direction. By using a position sensitive detector (PSD), which can measure both arrival time and transverse position of the charged particles, and assuming uniform magnetic and electric fields, the full three dimensional momentum information can be calculated from simple classical equations \cite{GisselbrechtRSI2005}.

If the electrons hit the PSD at an integer multiple of the cyclotron period, the transverse position of these electrons is independent of their initial transverse momentum. The corresponding points in the time of flight (ToF) spectrum are called magnetic nodes. At these nodes the momentum information is no longer unambiguous, leading to loss of data \cite{DornerPhysRep2000,Ullrich03rpp}. In the CIEL design, the time-of-flight of all electrons falls between two adjacent magnetic nodes, allowing for $4\pi$-angle resolved measurements without data loss.

The spectrometer was designed to be compact. The dimensions of the extraction region on both the electron and and ion detector sides as well as the length of the drift tube are shown in Figure~\ref{Fig:CIEL}A. The extraction fields are on the order of 5-15~\si{\volt/\cm}, and the detected range of electron kinetic energies is defined by adjusting the external magnetic field. Figure \ref{Fig:CIEL}B introduces the spherical coordinate system used for analysis of measurement data: the elevation $\theta$ defines the angle with respect to the light (linear) polarization direction, and the azimutal angle $\varphi$ determines the position of the projection onto the plane defined by the laser propagation direction (x) and the spectrometer axis (y). 

The gas sample is delivered via a long needle with a length-diameter ratio of approximately 1000, resulting in a directional and confined effusive jet \cite{SteckelmacherJPD1978}. The needle is mounted on a manual three dimensional translation stage which allows for precise positioning of the gas target with respect to the XUV and IR focus, and can be biased to match the potential of the extraction field.  

The detection system of the spectrometer consists of two commercially available PSDs (RoentDek Handels GmbH) based on multi-channel plates and delay line detectors, installed on both sides of the spectrometer, i.e. one to detect ions and the other electrons. 
The ion anode has a standard design \cite{JagutzkiNIMA2002}, while the electron detector provides unambiguous time and position information for multiple hits \cite{JagutzkiIEEETNS2002}. This gives us the possibility to detect one ion in coincidence with several correlated electrons.

% ----------------------------------------
\section{XUV-IR Interferometry in the Few Attosecond Pulse Regime}

The combination of the CEP-controlled, few-cycle laser source, high repetition rate, efficient HHG and a highly-stable XUV-IR pump-probe interferometer together with a 3D photoelectron/ion spectrometer that can record several correlated charged particles in coincidence establish excellent experimental conditions for exploring the physics of the interaction of few pulse APTs with atomic and molecular  systems. 

Recently, we studied single photoionization of helium by a few attosecond pulses (two or three pulses) in combination with a weak infrared (dressing) laser field~\cite{ChengPNAS2020}. Instead of using the XUV-IR interferometer described above, the IR laser pulses used for the generation were not eliminated by a metallic filter, but propagated collinearly with the APTs, after attenuation by an aperture, to the interaction region of the 3D spectrometer. Both the IR intensity and the XUV-IR delay were therefore fixed. The recorded photoelectron spectra were found to differ considerably depending on whether two or three attosecond pulses were used for the ionization. In the case of two pulses, the spectra were similar to those obtained with XUV-only radiation, except for a shift of the photoelectron peaks induced by the dressing IR field, towards high or low energy depending on the electron emission direction (up or down, relative to the laser polarization direction). In the case of three pulses, additional peaks, so-called sidebands, appeared exactly in between the peaks due to XUV-only absorption. They can be attributed to a two-photon process, where a harmonic photon is absorbed and an additional IR photon is either absorbed or emitted \cite{PaulScience2001}. The reader is referred to our earlier work~\cite{ChengPNAS2020} for details on the experiment and the simulations and for an intuitive interpretation of the results in terms of attosecond time-slit interferences. 

\begin{figure}[ht]
\includegraphics[width=1\linewidth]{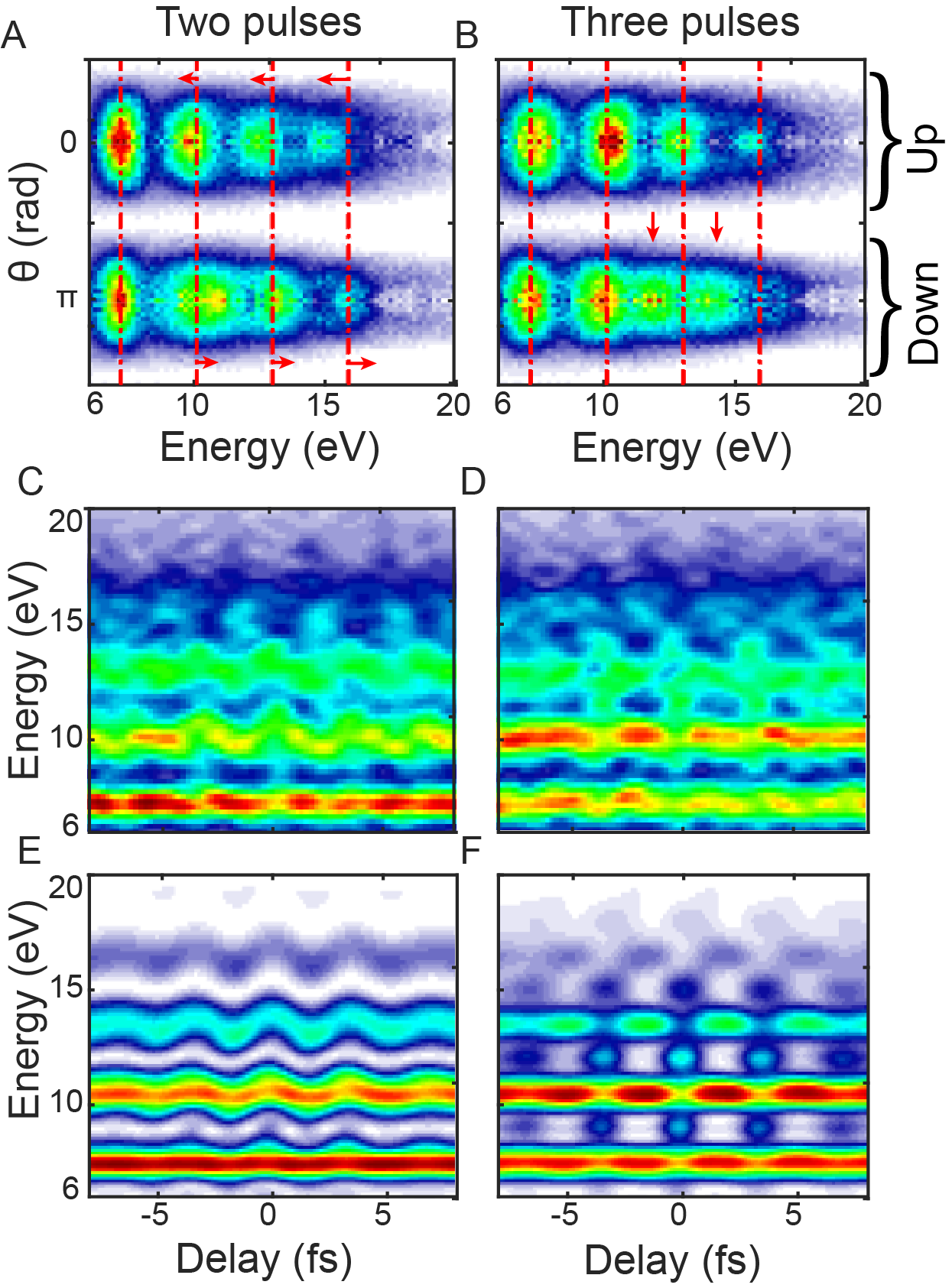}%
\caption{\textbf{Two-color photoionization of helium.} (A,B) Angular-resolved spectograms. Red dashed lines mark the position of absorption peaks corresponding to odd harmonics in the XUV only case. (C,D)  Experimental and (E,F) simulated XUV-IR delay traces, by integrating in the down direction. A, C, and E show the two pulse-case (CEP $\pi$/2), and B, D, and F the three pulse-case (CEP 0).  \label{Fig:Delay}}%
\end{figure}

In the present work, we use the XUV-IR pump-probe interferometer to record photoelectron spectra in helium as a function of XUV-IR delay in the  two- and three attosecond pulse cases. Figure~\ref{Fig:Delay} (A,B) present experimental photoelectron spectra at zero delay after integration over the azimutal angle $\varphi$ as a function of the elevation angle $\theta$, while Figure~\ref{Fig:Delay} (C,D) and (E,F) show spectra integrated over $2\pi$ solid angle in the down direction as a function of delay for the two- and three pulses cases, respectively. (C,D) are experimental results while (E,F) are simulations using the Strong-Field Approximation \cite{LewensteinPRA1994,QuereJMO2005,ChengPNAS2020}, with an IR intensity of \SI{6e10}{\watt/\centi\meter^2}.  The simulations agree well with the experimental measurements and illustrate the main features of the results, without experimental noise and/or irregularities. 

Figure~\ref{Fig:Delay} (A,B) reproduce and confirm our previous results \cite{ChengPNAS2020}, with energy-shifted photoelectron spectra in the two pulse case and the apparition of sidebands in the three-pulse case in the down direction. The energy shift can intuitively be understood by a classical picture, as in streaking \cite{KienbergerScience2002}. An electron emitted due to ionization by an attosecond pulse with a momentum $\textbf{p}$ gains or looses momentum from the IR field: $\textbf{p}\rightarrow \textbf{p} - e\textbf{A}(t_i)/c$, where $\textbf{A}(t_i)$ is the vector potential at the time of ionization $t_i$, $e$ the electron charge and $c$ the speed of light. The 3D momentum distribution becomes therefore asymmetric in the up- and down emission directions (Figure~\ref{Fig:Delay}A). In the three pulse case, the interpretation is more subtle. 
Here, in the down direction, sidebands appear between the peaks observed in the case of XUV-only radiation.  Depending on the delay between the XUV and IR fields, sidebands are observed in only one direction (here the down direction).

In Figure~\ref{Fig:Delay} (C,E), the photoelectron peaks are shifted towards lower or higher energies depending on the delay between the APTs and the IR pulses, similarly to attosecond streaking \cite{KienbergerScience2002,PazourekRMP2015}. Unlike a usual streaking trace, where a continuous photoelectron spectrum corresponding to ionization by a single attosecond pulse is periodically shifted in kinetic energy, here, the spectrum is in addition modulated by interference of the two electron wave packets created by the two attosecond pulses. The interference structure of the photoelectron spectrum provides spectral resolution, which is missing in ordinary streaking traces by a single attosecond pulse and may only be obtained by iterative retrieval procedures.

Figure~\ref{Fig:Delay} (D,F) shows simulated and experimental photoelectron spectra vs. delay for the three pulse case.
The results are quite different from those obtained with two attosecond pulses. Instead of modulations of the kinetic energy, sidebands appear at certain delays, similarly to RABBIT spectrograms \cite{Lopez-MartensPRL2005}, but with a major difference: The sidebands observed in the present case oscillate with a periodicity equal to the laser period, and not twice the laser period as in RABBIT.  Oscillating sidebands are also observed in the up direction, but with a phase shift of $\pi$ compared to the down direction. The asymmetry in the up- and down directions can be attributed to a parity mixing effect and is only observed in the few attosecond pulse case \cite{ChengPNAS2020}. 

The photoionization experiments presented here combining few attosecond pulses and a weak infrared field have similarities with streaking (for the two pulse case) and RABBIT (for the three pulse case) experiments, but are also distinctly different, presenting both challenges and new possibilities, especially concerning spectral resolution. In addition, the high repetition rate enables increased statistics, which is essential for full three-dimensional momentum detection. Advanced detection modes like coincidence become possible. Such a case is discussed in the next section.

% ---------------------------------------- 
\section{Single-Photon Double Ionization } 

Single-photon double-ionization in atoms and molecules is one of the most fundamental processes which leads to the emission of correlated photoelectron pairs \cite{ChandraJPB2002,ChandraPRA2004,AkouryScience2007}. In the simplest two-electron system, helium, previous works carried out primarily at synchrotron facilities have measured the absolute Triply Differential Cross Section (TDCS) for a range of excess energies and angular configurations \cite{BriggsJPB1999, AvaldiJPB2005}. The availability of attosecond techniques to probe single-photon double ionization \cite{ManssonNP2014} opens up new prospects for measuring the evolution of electron correlation in time. 
  
In order to demonstrate the unique capabilities provided by our high repetition-rate setup, we report the first ever results on single-photon double ionization of helium with full 3D momentum detection using an attosecond light source. Taking benefit of the full imaging capabilities of the CIEL spectrometer, our measurements show that processes hitherto unexplored by attosecond science are now within reach.   

Measuring double-ionization in helium presents, however, a few challenges for traditional attosecond light sources. Firstly, the threshold for this process at 79\,eV is beyond the energies easily achieved with standard HHG in argon using near-IR pulses. This can be circumvented by a change of generation gas, and as shown in section 2.4, we can reach sufficiently high energies by using neon.  Secondly, to perform a kinematically complete experiment we need to detect several charged particles in coincidence. We are thus limited in the final acquisition rate to one tenth of the repetition rate, to ensure that the event rate is below one event per shot, i.e. to minimize the likelihood of false coincidences. To reduce the number of single ionization events by absorption of low-order harmonics, leading to photoelectrons with similar energies as those due to double ionization by high-order harmonics, we use a zirconium filter which only transmits photon energies at $\approx$65\,eV and above.

\begin{figure}[ht!]
\includegraphics[width=1\linewidth]{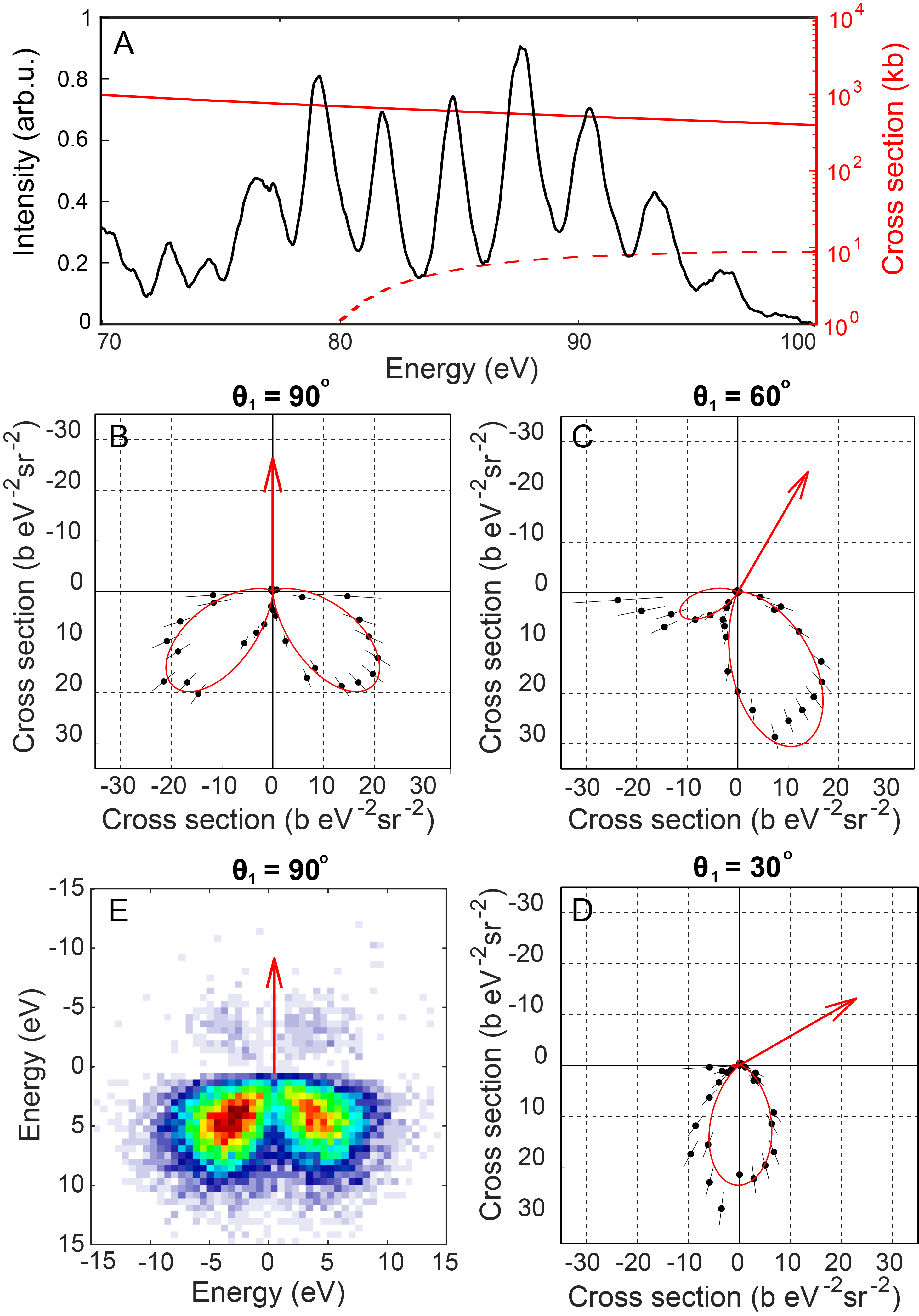}%
\caption{\textbf{Single-photon double ionization of helium.} (A) XUV spectrum (black) and cross section for single-photon single (solid red) and double (dashed red) ionization of helium \cite{SamsonJoESaRP2002,SamsonPRA1998}.  (B-D) TDCS for second electron (black dots) with equal energy sharing and co-planar emission, with emission angle of the first electron being \SI{90}{\degree}, \SI{60}{\degree}, and \SI{30}{\degree} (red arrow), compared to Eq. \ref{Eq:Wannier} (red curve). (E) Distribution of kinetic energies of second electron, integrated over all azimutal angles, when the first electron is emitted at $\theta_1 =$ 90\si{\degree} for co-planar and equal energy sharing emission.\label{Fig:CS}}%
\end{figure}

Figure \ref{Fig:CS}A shows a typical XUV spectrum generated with neon after this filter (black), along with the cross sections of both the single (red curve) and double (dashed red line) ionization process. In the photon energy region of 79-100\,eV, the average ratio of the double versus single ionization cross section is $\sim$1\%. With a coincidence rate of 20\,kHz, we expect a maximum possible acquisition rate of double ionization events to be 200\,Hz. Accounting for the total detection efficiency of typically $\sim$15\% for triple coincidence (two electrons and the doubly charged ion), the maximal acquisition rate for a 200\,kHz system reduces further to 30\,Hz, which is comparable to acquisition rates achievable at synchrotron facilities. 

To compensate for the relatively low XUV photon flux in this energy range, e.g. compared to the single photoionization experiments presented in Section 3, the backing pressure of the effusive jet for gas delivery in our spectrometer is increased and the jet is moved very close to the focus. 
We achieved a final detection rate of $\sim$\SI{15}{\hertz}, slightly lower than the nominal 30\,Hz, in order to operate the light source with good long-term stability over 40 hours. Clearly, without a high laser repetition rate, this experiment would be challenging.  

Figure \ref{Fig:CS}B-D shows the measured TDCS for equal energy sharing of the two electrons (E$_1$=E$_2$=5$\pm$1 eV)  with emission angles of the first electron of $\theta_1 =$ \SI{90}{\degree}, \SI{60}{\degree}, and \SI{30}{\degree} with respect to the polarization axis, and with a total kinetic energy E$_1 + $E$_2  = 10 \pm 1.5$ \si{\electronvolt}.  The normalization is done according to the procedure outlined in Bräuning et al. \cite{BruningJPB1998} using a value of the total cross section of 7.23\,kb at 10\,eV above threshold \cite{SamsonPRA1998}. The exact analytical derivation of the TDCS at threshold \cite{HuetzJPB1991,BriggsJPB1999} allows us to further analyse our data. In helium, for a two-electron state with $^1$P$^\mathrm{o}$ symmetry, the differential cross-section for equal energy sharing can be written as \cite{WannierPR1957,HuetzJPB1991,BriggsJPB1999}:
\begin{equation}
        \frac{d^3\sigma}{dE_1 d\Omega_1 d\Omega_2} = 
         a_g(E_1,E_2,\theta_{12})\left( \cos{\theta_1}+\cos{\theta_2}\right)^2, \label{Eq:Wannier}
\end{equation}
where $\theta_1$ and $\theta_2$ are the emission angles of the two electrons with respect to the polarization axis (in accordance with the coordinate system defined Figure \ref{Fig:CIEL}B). The term  $\left(\cos{\theta_1}+\cos{\theta_2}\right)$, arises from the geometry of the light-matter interaction. A quantum-mechanical description of double photo-ionization with an electron-pair in the continuum must obey the Pauli principle and symmetries depending on the total orbital angular momentum, $L$, the total spin, $S$, and the parity of the final two-electron state. This leads to a selection rule prohibiting the back-to-back emission, that is observed as a node in the TDCS (see Figure \ref{Fig:CS}B-D). 
The complex amplitude, $a_g$, describes the correlation dynamics of the electron-pair and only depends on the excess energy and the mutual angle $\theta_{12}$. A Gaussian ansatz provides an excellent parametrization of this amplitude \cite{BriggsJPB1999,AvaldiJPB2005}: 
\begin{equation}
 a_g(E_1,E_2,\theta_{12})=a\exp\left(-4\ln{2}\left[(\theta_{12}-180^o)/\gamma\right]^2\right),
\end{equation}
where $a$ is a scaling factor depending on the cross section, and $\gamma$ is the full-width at half maximum correlation factor, which depends on the excess energy. This term affects the opening angle between the two "lobes" seen for example in Figure \ref{Fig:CS}B. A theoretical calculation \cite{MalegatPRA1999,HuetzPRL2000} predicts that the opening angle $\gamma$ should be around \SI{93}{\degree}. We find experimentally a value of $\gamma =$ \SI{90}{\degree}$\pm$\SI{3}{\degree}, in  better agreement than previous reported values of approximately \SI{85}{\degree} at this excess energy \cite{DornerPRA1998,AvaldiJPB2005}.  

The results in Figure \ref{Fig:CS}B-D were achieved by integrating over only a small total energy interval for comparison with previous works. However, since the ionizing XUV radiation has a very broad spectrum, this filtering does not give an complete representation of the two-electron wave-packet (EWP) dynamics. To visualize this EWP, we show in Figure \ref{Fig:CS}E the result obtained over the whole energy range, integrating over the azimutal angle and for a \SI{90}{\degree} emission angle of the first electron, as in Figure \ref{Fig:CS}B. Interestingly the variation in emission direction and energy of the second electron exhibits the nodal properties (selection rule) that are usually observed in the fully differential cross section. 
Using the pump-probe capabilities of our setup in a future experiment, the time evolution of the EWP can be studied by observing the change of the final state of the two-electron EWP in the continuum.

\section{Conclusion}
In this work, we present a compact, high-repetition-rate, attosecond light source and demonstrate its capability to perform time-resolved measurements and coincidence experiments. 
The high repetition-rate, enabled by the partially fiber-based laser architecture, opens up for applying the extraordinary time-resolution promised by attosecond science to new processes and phenomena.

We study single photoionization of He atoms using APTs consisting of only a few pulses in combination with a weak, delayed, IR field. This new regime opens up new exciting possibilities for control of photoemission through a tailored sequence of pulses. The case of two attosecond pulses, in particular, is promising, as the results can intuitively be understood as in streaking experiments. Compared to streaking with a single attosecond pulse, it adds spectral resolution due to the interference structure induced by the two attosecond pulses, similar to Ramsey spectroscopy, and lower intensity can be used for the dressing field because of the gained spectral resolution, thus less perturbing the system under study.
Generally, the potential of the few pulse attosecond train regime seems to have been widely overlooked by the attosecond science community that traditionally has either worked with single attosecond pulses (streaking) or long attosecond pulse trains (RABBIT).  

Our setup includes the possibility to add a second end-station in order to perform time-resolved surface science experiments (Fig.~\ref{Fig:beamline}). As in coincidence spectroscopy, these experiments strongly benefit from high repetition rate in order to avoid space charge related blurring effects in spectroscopy and imaging applications \cite{Chew2015}.
As the setup is designed for the two end-stations to be used simultaneously, the gas phase experiments can serve as a benchmark for simultaneous time-resolved studies on nanostructured surfaces. Our ultimate goal is to investigate and control charge carrier dynamics on the nanoscale with attosecond temporal resolution.

Finally, as a proof of principle, we successfully measure single-photon double ionization in helium by our broadband attosecond XUV source over a complete $4\pi$ solid angle. The next experiment will be to add a weak IR laser field in order to study the time evolution of the two-electron wave packet. More generally, our high-repetition rate setup is well adapted for the study of highly correlated many-body processes in the temporal domain. 

\begin{acknowledgements}
The authors acknowledge support from the Swedish Research Council and the Knut and Alice Wallenberg Foundation.
\end{acknowledgements}

% Create the reference section using BibTeX:
\bibliography{reflib_renamed.bib}
\bibliographystyle{nphotonics}

\end{document}